# Qué sabe el consumidor del daño ambiental provocado por el vaso desechable y la necesidad de sustituirlo

*What does the consumer know about the environmental damage caused by the disposable cup and the need to replace it*

*O que o consumidor sabe sobre os danos ambientais causados pelo copo descartável e a necessidade de substituí-lo*

**Alejandro Arturo Rivera Sánchez**
Tecnológico Nacional de México, Instituto Tecnológico de Ciudad Guzmán, México
aalexius@hotmail.com
https://orcid.org/0000-0001-5895-0226

**Guillermo José Navarro del Toro**
Universidad de Guadalajara, Centro Universitario de los Altos, México
navarromemo@hotmail.com
https://orcid.org/0000-0002-4316-879X

**Resumen**

El objetivo del presente trabajo fue conocer la cantidad y la frecuencia con que las personas de Arandas en la región de los Altos de Jalisco emplean los vasos desechables para luego saber qué tan dispuestas están en utilizar vasos comestibles elaborados con gelatina natural. Al respecto, vale comentar que estos no solo pueden ser nutritivos para quienes los consuman (pues la gelatina es un nutriente fortificante creada a partir de la piel y el hueso de cerdos y vacas), sino que también se podrían degradar en pocos días o ser ingeridos por los animales. Para recabar la información se empleó una encuesta constituida por seis interrogantes, la cual fue aplicada a 31 personas por vía telefónica y a otras 345 de manera personal (en ambos casos se aplicaron a jóvenes y adultos). Los resultados demuestran que los pobladores de esa localidad usan considerablemente vasos de plásticos en los distintos eventos que se realizan cada semana, los cuales son más numerosos en las fiestas patronales o de fin de año. Aun así, estas personas estarían dispuestas a cambiar estos hábitos, aunque para ello se deben tomar medidas que no afecten a las empresas de esa zona, las cuales trabajan principalmente con plásticos y generan un alto porcentajes de empleos.

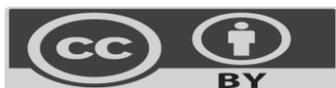

**Palabras clave:** contaminación, desechables, gelatina, medio ambiente, plástico, unicel.

## Abstract

The objective of this work was to know the amount and frequency with which the people of Arandas in the Altos de Jalisco region use disposable cups and then know how willing they are to use edible cups made with natural gelatin. In this regard, it is worth commenting that these can not only be nutritious for those who consume them (since gelatin is a fortifying nutrient created from the skin and bone of pigs and cows), but they could also be degraded in a few days or be ingested by animals. To collect the information, a survey consisting of six questions was used, which was applied to 31 people by telephone and another 345 personally (in both cases they were applied to young people and adults). The results show that the residents of that town considerably use plastic cups in the different events that take place each week, which are more numerous during the patron saint festivities or at the end of the year. Even so, these people would be willing to change these habits, although for this, measures must be taken that do not affect the companies in that area, which work mainly with plastics and generate a high percentage of jobs.

## Resumo

O objetivo deste trabalho foi conhecer a quantidade e a frequência com que os arandas da região dos Altos de Jalisco usam os copos descartáveis e a disposição para usar os copos comestíveis feitos com gelatina natural. Nesse sentido, vale comentar que estes não só podem ser nutritivos para quem os consome (já que a gelatina é um nutriente fortificante criado a partir da pele e do osso de porcos e vacas), mas também podem se degradar em poucos dias ou ser. ingerido por animais. Para a recolha das informações, foi utilizado um inquérito composto por seis questões, que foi aplicado a 31 pessoas por telefone e outras 345 pessoalmente (em ambos os casos foram aplicados a jovens e adultos). Os resultados mostram que os moradores daquela cidade usam bastante os copos plásticos nos diversos eventos que acontecem a cada semana, sendo mais numerosos durante as festas da padroeira ou no final do ano. Mesmo assim, essas pessoas estariam dispostas a mudar esses hábitos, mas para isso devem ser tomadas medidas que não afetem as empresas da área, que trabalham principalmente com plásticos e geram um alto percentual de empregos.

## Introducción

Cuando se inició la elaboración de productos derivados del petróleo, se buscó facilitar las actividades humanas. Sin embargo, nunca se previó que esa iniciativa causara efectos secundarios tan graves ni que llegaran a poner en riesgo la vida del planeta. Muchos de esos productos han sido ampliamente aceptados y están presentes en todas las actividades realizadas por el ser humano. Desgraciadamente, al ser desechados llegan a lagos, ríos y mares, y afectan los ecosistemas acuáticos y sus especies. Los residuos plásticos (bolsas, popotes, botellas, etc.) (Dirección General de Servicios de Documentación, Información y Análisis, 2019) arrojados al mar han formado una isla flotante de aproximadamente 1.6 millones de km$^2$, es decir, de tres veces el tamaño de Francia (Lebreton, 2019). De hecho, en tierra es común encontrar conjuntamente desechos orgánicos mezclados con bolsas, popotes, contenedores desechables domésticos e industriales, muebles, molduras, entre otros. Además, y debido a que no son biodegradables, se han convertido en un problema mundial directamente relacionado con el calentamiento global y el deterioro del medio ambiente y los ecosistemas.

Por otro lado, muchas empresas manufactureras de desechables que usan como materia prima derivados del petróleo se han ido instalando en poblaciones cada vez más pequeñas, de ahí que esas localidades dependan de ellas. Asimismo, es necesario decir que por el costo de esos productos sus pobladores son excelentes consumidores. Por ese extendido uso, cada vez que utilizan algún producto (vaso, envase, moldura, etc.), lo tiran al suelo o lo colocan en bolsas mezclados con desechos orgánicos caseros, terminan en los vertederos a cielo abierto de las poblaciones. Esto sucede porque al consumidor jamás se le explica la forma apropiada para deshacerse de los plásticos que han sido utilizados, lo que ha ocasionado grandes problemas en el planeta que son difíciles de combatir y aún más de revertir.

Si bien el humano siempre ha afectado los ecosistemas, con la llegada del desechable ese proceso se aceleró, lo que se evidencia en el surgimiento de nuevos eventos naturales, como los "fenómenos del niño y de la niña" (Vásquez, Dorado y Rosales, 1996), la ausencia o abundancia inexplicable de lluvias o la marea roja. Todo lo anterior ha aumentado el peligro de extinción de muchas especies, así como el incremento de la temperatura, la emisión de gases a la atmosfera por procesos contaminantes de industrias y vehículos, la "basura orgánica" que produce cantidades enormes de metano y la "basura inorgánica" de difícil degradación.

Esos contaminantes, lógicamente, se pueden ver en zonas rurales, campos, ríos, playas, mares, zonas urbanas, calles, vertederos, sistemas de desagüe y mantos freáticos, que son saturados por los lixiviados que escapan de botes plásticos y de cartón plastificado. Para revertir estos efectos se debe educar a las nuevas generaciones, las cuales deben cambiar sus hábitos en cuanto a clasificar y separar la basura, reciclarla y seleccionar productos de limpieza y de transporte que sean amigables con el medio ambiente. Asimismo, se deben idear proyectos que permitan encontrar nuevas materias primas para sustituir los derivados del petróleo.

Por ello, el objetivo del presente trabajo fue conocer la cantidad y la frecuencia con que las personas de Arandas en la región de los Altos de Jalisco emplean los vasos desechables para luego saber qué tan dispuestas están en utilizar vasos comestibles elaborados con gelatina natural. Al respecto, vale comentar que estos no solo pueden ser nutritivos para quienes los consuman (pues la gelatina es un nutriente fortificante creada a partir de la piel y el hueso de cerdos y vacas) (*OKDiario*, 12 de diciembre de 2016), sino que también se podrían degradar en pocos días o ser ingeridos por los animales.

## Marco teórico

Los productos que usan derivados del petróleo como materia prima para su elaboración son extremadamente comunes en la vida de los seres humanos (oficina, taller, casa, vestido, etc.). Estos fueron creados en principio para facilitar diversas actividades, aunque muchos se usan solo una vez, por lo que rápidamente se convierten en basura no reutilizable. Los acuerdos internacionales promovidos por la Organización para la Cooperación y Desarrollo Económico (OCDE) (Bremer, 2018) para evitar el uso de materiales contaminantes del medio ambiente han sido firmados por México, lo que ha obligado a la nación a promulgar leyes para mitigar los efectos secundarios de esos productos (Dirección General de Servicios de Documentación, Información y Análisis, 2018). Esas leyes prohíben el uso de materia prima derivada del petróleo para la fabricación de empaques de comida de unicel, popotes y bolsas plásticas, de ahí que se deba optar por materiales biodegradables.

Por otro lado, la OCDE establece que en México —del total de residuos que se generan— solo 9.6 % son reciclados, lo que debe a la baja conciencia de la gente para cuidar y mejorar el medio ambiente (Muñoz, 2019). Por ese motivo, con el presente proyecto se procura generar conciencia sobre esta realidad en habitantes de la población de Arandas en la región de los Altos de Jalisco, ya que su economía está basada en la elaboración de productos desechables, los cuales son distribuidos a otros lugares del país e incluso a algunas naciones de Centro América.

En efecto, si bien Arandas se caracteriza por su producción de tequila, lácteos, carnitas y muebles, también es cierto que esa localidad actualmente cuenta con 20 empresas manufactureras de productos de plástico, las cuales han estado en la región por más de 30 años y en la actualidad generan 6000 empleos directos y 2000 indirectos. De acuerdo con el presidente de Asociación Nacional de Industrias del Plástico (ANIPAC), José Anguiano, 90 % de las fábricas productoras de plástico tienen la capacidad y la infraestructura para producir plástico reciclado como materia prima, pero los productos que sirven para contenedores de alimentos y bebidas para consumo humano tienen que ser elaborados con materia prima virgen (Ramírez Gallo, 25 de agosto de 2018).

### Jalisco en la búsqueda de prohibir el uso de recipientes plásticos

En el año 2018, el Congreso de Jalisco se reunió con el propósito de analizar y aprobar una iniciativa de ley para prohibir productos desechables elaborados con unicel o plástico. Se tomó en consideración el hecho de que el estado es uno de los principales productores de plástico y recipientes desechables en el país. De hecho, se previó que si se aprobaba la iniciativa presentada por el regidor Juan José Cuevas en el municipio de Puerto Vallarta (y de manera posterior las diputadas de Movimiento Ciudadano, Verónica Jiménez y Lourdes Martínez), se propondrían reformas a la Ley Estatal del Equilibrio Ecológico y de Protección al Ambiente, y a la Ley de Gestión Integral de Residuos. Esto, sin embargo, impactaría directamente en unos 40000 empleos del sector, quienes podrían ver en riesgo sus fuentes de trabajo.

Por tal motivo, el análisis económico es sumamente importante, ya que con estas leyes se afecta a empresarios, trabajadores y familias. En este tipo de análisis, por tanto, se debe de considerar que con la prohibición las empresas verían afectados sus contratos para la adquisición de insumos, maquinaria y hacer frente a sus deudas, lo que provocaría la quiebra, como lo señaló Arturo Álvarez, presidente de la sección Plásticos de la Cámara Regional de la Industria de la Transformación (Careintra). Según Careintra, el municipio de Arandas en la zona de los Altos es donde más plásticos se producen a escala nacional, mientras que Jalisco es el principal exportador de envases, empaques, embalaje y productos desechables de plástico para la región del Pacífico (Romo, 31 de julio de 2018).

El titular de la Secretaría de Medio Ambiente y Desarrollo Territorial, Sergio Humberto Graf Montero, avaló la medida, pero pidió que antes se elaborara un diagnóstico y un plan de acción. Para ello, se podría empezar por los popotes y luego continuar con las bolsas y desechables plásticos usados en ciertos lugares (como los supermercados) (Romo, 31 de julio de 2018).

Vale destacar que con esas leyes no se busca frenar la actividad económica, sino propiciar la generación de productos empleando otras alternativas ecológicas (Blanco, 13 de julio de 2018). El objetivo es procurar que los empresarios no abandonen el estado o sus actividades económicas, sino que se conviertan en los primeros industriales que inviertan en la transformación de sus equipos para adaptarlos a las nuevas exigencias.

Una opción, por tanto, sería emplear materias primas que servirían para producir "vasos comestibles", ya que los actuales no pueden ser reutilizados. En principio, la propuesta se enfocaría en la fabricación de vasos de gelatina, los cuales pueden ser comestibles o usados como composta para abono altamente orgánico.

**Disposición**

Uno de los antecedentes más importantes con respecto a las disposiciones sobre el uso de bolsas de plástico proviene del municipio de Pachuca (Mota, 9 de marzo de 2018), donde el Reglamento de Protección Ambiental y Cambio Climático prohíbe que los establecimientos otorguen bolsas de plástico para acarreo de productos en los comercios. La delegación de Pachuca de la Cámara Nacional de Comercio, Servicios y Turismo (Canaco Servytur) tiene registrados 20000 establecimientos a los cuales se les comunicó que deben cumplir esta disposición.

Por su parte, en Jalisco se hicieron iniciativas de ley para prohibir el uso de plásticos en el estado a partir del 2020. Ante el Congreso del Estado, y debido a las toneladas de peces y aves que mueren diariamente víctimas de la contaminación generada por plástico y polietileno —principalmente en las zonas turísticas de Puerto Vallarta—, las diputadas Verónica Magdalena Jiménez Vázquez y María Lourdes Martínez Pizano presentarán una iniciativa para *prohibir de manera gradual* el uso y fabricación de los productos elaborados con plástico y unicel, ambos derivados del petróleo.

En rueda de prensa, las diputadas en compañía del regidor del municipio de Puerto Vallarta, Juan José Cuevas García, anunciaron que en sesión plenaria presentarían la iniciativa porque el uso de plástico ha estado ocasionando daños irreparables al planeta y el problema se agudiza, ya que anualmente se producen más de 300 millones de toneladas de plástico que terminan contaminando cuerpos lacustres, principalmente ríos y mares.

Con esta propuesta, se pretende reformar la Ley Estatal de Equilibrio Ecológico y la Protección al Ambiente y la Ley de Gestión Integral de Residuos con el fin de que se aplique a los 125 municipios del estado. La iniciativa plantea cambiar el uso de plásticos por sustitutos reutilizables, así como fomentar el uso de envases y recipientes adecuados para evitar comprar bebidas y comida en plásticos, unicel o bolsa; además, disminuir la compra de agua embotellada e incentivar el uso de filtros de agua y mejorar la gestión del reciclaje desde los comercios y hogares.

Asimismo, se espera que los prestadores de servicios de playas concienticen a productores y usuarios de cambiar el plástico y unicel por materia prima como la gelatina; con ello, se dará el primer paso para su uso, y posteriormente se buscará y demandará este tipo de productos en sus lugares de origen.

Cabe destacar que se esperaba que esta iniciativa entrara en vigor a partir del primero de enero del 2020. Además, los legisladores expusieron que a los industriales se les brindarán oportunidades para que elaboren sus productos "comestibles" utilizando materia prima biodegradable, como fécula del maíz.

## Procesamiento de la información

Para recabar la información se empleó una encuesta constituida por seis interrogantes, la cual fue aplicada a 31 personas por vía telefónica y a otras 345 de manera personal (en ambos casos se aplicaron a jóvenes y adultos).

La primera pregunta (en ambas encuestas) permitió establecer la cantidad de vasos que usan por semana los encuestados, ya que de ellos depende la acumulación de la basura no degradable (tabla 1).

**Tabla 1.** Respuestas a la pregunta número 1 (cantidad de vasos usados por semana)

| *# vasos* | *# personas (encuesta 1)* | *# personas (encuesta 2)* |
|---|---|---|
| **0** | **7** | **19** |
| **1-3** | **16** | **150** |
| **4-10** | **7** | **26** |
| **Más de 10** | **1** | **127** |

Fuente: Elaboración propia

Tomando en consideración las cifras ofrecidas en la tabla 1, y si se toma en cuenta que Arandas tiene una población de 84 966 habitantes "según una estimación poblacional de la Encuesta Intercensal para 2019" (IIEG, 2019), se podría decir que al año en esa localidad se podrían estar usando un promedio de entre 13254696 y 2´2091160 vasos desechables plásticos. Ahora bien, según el último informe de PlasticsEurope, Plastics the facts 2018, la producción mundial de plásticos en 2017 alcanzó los 348 millones de toneladas (*MundoPLAST*, 2017). Eso significa que la región de Arandas estaría participando con el .0051 % de la producción mundial de plásticos en 2017, solo tomando en consideración el consumo de vasos plásticos desechables.

Por otra parte, en la segunda interrogante planteada a las personas se les consultó sobre el tipo de material con que están elaborados los vasos que usan cada semana en distintos eventos (tabla 2).

**Tabla 2.** Respuestas a la pregunta número 2 (material de los vasos usados por semana en eventos)

| *Tipo de producto* | *# personas (encuesta 1)* | *# personas (encuesta 2)* |
|---|---|---|
| **Desechables de plástico** | **20** | **289** |
| **Vidrio** | **4** | **20** |
| **Plástico** | **7** | **36** |

Fuente: Elaboración propia

Los datos de la tabla 2 indican que la mayoría de la población de Arandas usa vasos plásticos desechables para realizar sus eventos. Al respecto, vale acotar que en esta localidad se realizan entre 5 y 10 eventos masivos por semana, a los cuales asisten más de 100 personas. Estas cifras, sin embargo, pueden aumentar significativamente en los meses en que, por ejemplo, se efectúan las fiestas patronales o de fin de año.

En la pregunta número 3 se buscó conocer si sabían o no sobre el daño que provocan al medio ambiente los plásticos que se arrojan en la basura. Los resultados fueron bastante alentadores, ya que en la tabla 3 se muestra que una gran cantidad de gente sabe lo que se ocasiona con el uso de recipientes plásticos.

**Tabla 3.** Respuestas a la pregunta número 3 (conocimiento sobre la contaminación que provocan los plásticos)

| *Respuesta* | *# personas (encuesta 1)* | *# personas (encuesta 2)* |
|---|---|---|
| **Sí** | **29** | **327** |
| **No** | **2** | **18** |

Fuente: Elaboración propia

En la tabla 3 se puede apreciar que las personas son conscientes del deterioro ambiental provocado por los desechos inorgánicos; el problema, sin embargo, radica en que aun sabiendo esta situación, pareciera que en la vida cotidiana no aplican ese conocimiento. Esto demuestra que son insuficientes las campañas que se hacen a través de los medios tradicionales o digitales de comunicación. Por tanto, es necesaria una contribución especial de la escuela, de las empresas que generan esos desechos y de la familia para cambiar esos hábitos de contaminación.

En la pregunta número 4 se planteó a los entrevistados si estarían dispuestos a usar de manera recurrente vasos comestibles elaborados con materiales amigables con el ambiente. En la tabla 4 se aprecia que la mayoría está dispuesta a realizar el cambio.

**Tabla 4.** Respuestas a la pregunta número 4 (disposición para utilizar productos que sustituyan a los vasos comestibles)

| *Respuesta* | *# personas (encuesta 1)* | *# personas (encuesta 2)* |
|---|---|---|
| Sí | 30 | 329 |
| No | 1 | 16 |

Fuente: Elaboración propia

Sobre la pregunta anterior, vale señalar que muchos productores de comestibles biodegradables usan materiales más amigables con el medio ambiente (*BBC Mundo*, 18 de diciembre de 2013), algunos de los cuales se mencionan a continuación:

- Plástico "cultivado" a partir de hongos: Material mezclado con residuos de agricultura.
- Vasos desechables de seda y camarones: Es un material biocinético que toma el diseño y componentes de la concha de camarón y las proteínas de la seda.
- Patatas para producir plástico: Básicamente es una resina compuesta que cuando se comprime a partir de calor y presión parece plástico.
- Gelatina que se extrae de huesos y piel de los animales.

En la pregunta número 5 se obtuvieron los datos más interesantes con respecto al número de personas que conocen el tiempo en que se puede degradar un vaso plástico desechable (tabla 5).

**Tabla 5.** Respuestas a la pregunta número 5 (conocimiento del tiempo en que tarda en degradarse un vaso plástico desechable)

| Tiempo para que se degrade el plástico | # personas (encuesta 1) | # personas (encuesta 2) |
| --- | --- | --- |
| No sé | 5 | 75 |
| Desconozco el dato | 2 | 53 |
| Depende del polímero | 1 | 2 |
| 1 año | 1 | 10 |
| 5 años | 1 | 1 |
| Más de 10 años | 1 | 1 |
| 50 años | 2 | 1 |
| 100 años | 6 | 1 |
| 150 años | 3 | 1 |
| 150-400 años | 1 | 1 |
| 250 años | 1 | 1 |
| 500 años | 2 | 98 |
| 1000 años | 3 | 5 |

**Nota:** En la encuesta telefónica contestaron 29 personas, y en la encuesta personal 250.

Fuente: Elaboración propia

En relación con la tabla anterior, se puede decir que Earthgonomic México A. C. (9 de enero de 2017) indica que los vasos de unicel, los vasos térmicos de cartón recubiertos de cerámica y los contenedores de plástico fabricados con PET tienen un promedio de vida para su degradación de aproximadamente 1000 años, mientras que materiales naturales hechos con la cáscara de plátano pueden hacer el mismo proceso en un tiempo de 5 a 10 días.

Finalmente, la pregunta número 6 se enfocó en conocer si las personas sabían cómo debían tratarse los desechos inorgánicos de difícil degradación como el plástico (tabla 6).

**Tabla 6.** Respuestas a la pregunta número 6 (conocimiento sobre el tratamiento de los vasos desechables)

| *Destino del vaso que se ha utilizado* | # personas (encuesta 1) | # personas (encuesta 2) |
|---|---|---|
| **Se colocan separados con todo el plástico** | 20 | 240 |
| **Se colocan en la basura mezclada con la basura orgánica** | 9 | 72 |
| **Se arrojan en cualquier lugar** | 2 | 33 |

Fuente: Elaboración propia

A partir de los datos recabados en la anterior interrogante, se puede decir, de forma general, que las generaciones más jóvenes que asisten a las universidades tienen más conciencia de la forma en que deben deshacerse los vasos de plásticos hechos con derivados del petróleo. Si bien es cierto que las zonas metropolitanas existe una regulación con respecto a la separación de basura por su tipo, la mayoría las personas que viven en otras zonas no realiza esta clasificación.

## Discusión

Uno de los principales problemas a nivel mundial es el alto impacto que tiene el uso de materiales hechos con plástico, en especial aquellos que se usan una sola vez. Esto queda en evidencia en la isla basura ubicada en el océano Pacífico, cuyo tamaño es tres veces el de Francia. Por este motivo, se debe emprender acciones que permitan cambiar esa realidad, para lo cual es esencial que se empiecen a utilizar productos biodegradables.

Si bien en muchas ciudades existen programas que promueven la separación de la basura según su tipo, la realidad demuestra que en la mayoría de las localidades no se siguen estas recomendaciones. Además, en muchos casos cuando los desechos que ya fueron clasificados por las personas llegan a los camiones de basura o a los vertederos, son nuevamente reagrupados sin seguir ningún criterio de diferenciación.

En síntesis, se puede afirmar que en nuestro país se debe reeducar a la gente, para lo cual se puede usar la estrategia de hacerles ver que muchos de esos desechos pueden generar ingresos debido a las posibilidades de ser reusados bien sea como materia prima de otros productos o como compostas o parte de alimento para animales (en el caso de los desechos biodegradables).

## Conclusiones

Debido al incremento de materiales contaminantes, se han establecido distintos programas ambientales. En la presente investigación se pudo establecer con mucha claridad que la gente está dispuesta a cambiar sus hábitos para adaptarse a las nuevas exigencias de un mundo más habitable. Para ello, una opción sería el uso de vasos comestibles hechos con gelatina, pero esta iniciativa debe ir acompañada de políticas y financiamientos que ayuden a las industrias a adoptar esos métodos.

De igual manera, se deben impulsar cursos y talleres para educar a la gente en torno al grave problema que están generando los desechos inorgánicos, lo cual probablemente ocasionará un daño que puede ser irreversible en pocos años. Con esa capacitación al público en general, se estarán cumpliendo con los programas que la ONU inició para concientizar a las naciones sobre las causas del cambio climático.

En este sentido, se tendrán que establecer convenios con supermercados, tiendas de conveniencia y abarrotes para que eviten vender productos elaborados de derivados del petróleo, y empiecen a sustituirlos por otros que sean más amigables con el medio ambiente.

Por supuesto, en el caso concreto de Arandas —que se localiza en la región de los Altos de Jalisco—, se debe de tomar en consideración que en esta localidad existe un gran número de plantas fabricantes de productos desechables, las cuales generan un gran porcentaje de empleos.

Por lo tanto, se debe hacer un trabajo extenuante que, a sugerencia de algunos entrevistados, tiene que ser liderado por especialistas de las universidades, ya que son ellos quienes tienen los conocimientos necesarios para revertir el daño ocasionado al medio ambiente.

## Referencias

| Rol de Contribución | Autor (es) |
|---|---|
| **Conceptualización** | (Guillermo José Navarro del Toro) |
| **Metodología** | (Alejandro Arturo Rivera Sánchez) |
| **Software** | (Alejandro Arturo Rivera Sánchez) |
| **Validación** | (Guillermo José Navarro del Toro) (Alejandro Arturo Rivera Sánchez) (Igual) |
| **Análisis Formal** | (Alejandro Arturo Rivera Sánchez) |
| **Investigación** | (Guillermo José Navarro del Toro) |
| **Recursos** | (Alejandro Arturo Rivera Sánchez) |
| **Curación de datos** | (Alejandro Arturo Rivera Sánchez) |
| **Escritura - Preparación del borrador original** | (Guillermo José Navarro del Toro) (Alejandro Arturo Rivera Sánchez) (Igual) |
| **Escritura - Revisión y edición** | (Guillermo José Navarro del Toro) (Alejandro Arturo Rivera Sánchez) (Igual) |
| **Visualización** | (Alejandro Arturo Rivera Sánchez) |
| **Supervisión** | (Guillermo José Navarro del Toro) |
| **Administración de Proyectos** | (Guillermo José Navarro del Toro) (Alejandro Arturo Rivera Sánchez) (Igual) |
| **Adquisición de fondos** | (Guillermo José Navarro del Toro) |